\documentclass[preprint]{aastex63}
\usepackage{amsmath}
\graphicspath{{./}{figures/}}
\begin{document}

\title{Analysis of Time-Distance Helioseismology for Detection of Emerging Active Regions}
\correspondingauthor{John T. Stefan}
\email{jts25@njit.edu}

\author[0000-0002-5519-8291]{John T. Stefan}
\affiliation{Department of Physics, New Jersey Institute of Technology, Newark, NJ 07102}

\author[0000-0003-0364-4883]{Alexander G. Kosovichev}
\affiliation{Department of Physics, New Jersey Institute of Technology, Newark, NJ 07102}
\affiliation{NASA Ames Research Center, Moffett Field, Mountain View, CA 94040}

\author[0000-0001-7483-3257]{Andrey M. Stejko}
\affiliation{Department of Physics, New Jersey Institute of Technology, Newark, NJ 07102}

\received{12/02/20}
\revised{02/01/21}
\revised{02/23/21}
\accepted{-}
\begin{abstract}

A time-distance helioseismic technique, similar to the one used by Ilonidis et al, is applied to two independent numerical models of subsurface sound-speed perturbations to determine the spatial resolution and accuracy of phase travel time shift measurements. The technique is also used to examine pre-emergence signatures of several active regions observed by the Michelson Doppler Imager (MDI) and the Helioseismic Magnetic Imager (HMI). In the context of similar measurements of quiet sun regions, three of the five studied active regions show strong phase travel time shifts several hours prior to emergence. These results form the basis of a discussion of noise in the derived phase travel time maps and possible criteria to distinguish between true and false positive detection of emerging flux.

\end{abstract}

\keywords{helioseismology -- solar active regions -- solar magnetic flux emergence}

\section{Introduction}

As society relies more heavily on space-based infrastructure, there is an increasing need for reliable prediction and forecasting of space weather and geomagnetic activity. It is well-understood by now that the Sun is the main influence on the geospace environment; therefore, solar variability must form the basis of any space weather forecasting tools. Decades of observation have shown that the Sun is subject to a 22-year magnetic cycle, with sunspot number and active region size, strength, and location being the main manifestations of this cycle. Active regions are characterized by their strong magnetic fields which contribute to their high solar flare productivity, relative to other areas of the Sun. While determining when an active region will flare remains a complex and open problem, forecasting the development of active regions is slightly more straightforward.

A well-regarded hypothesis for the development of active regions is the rising flux tube model. MHD modeling has shown that strong shearing, such as found at the boundary between the solar radiative and convective zones, can turn weak poloidal magnetic fields into strong toroidal fields \citep{Hughes_mag,Brummel_mag}. The resulting concentration of magnetic field, or flux tube, experiences a buoyant force which causes the tube to bend and rise to the surface \citep{Cline_mag}. This upside-down "u"-shaped tube explains the bipolar nature of many active regions \citep{Parker_bipole}. Importantly, the sub-surface origin of active regions should allow for detection before the magnetic flux reaches the solar surface.

There are many helioseismic techniques which are useful in performing this detection, some of which have already had success in identifying existing active regions on the far-side of the Sun. Far-side imaging is accomplished primarily using two different methods: acoustic holography and time-distance analysis. In acoustic holography, the far-side wavefield is obtained from near-side measurements and inhomegeneities in the wavefield, which are measured by the relative phase-shift, can be attributed to the strength of magnetic field \citep{hseisholo,Gizon_holo}. Time-distance analysis, as the name suggests, measures shifts in phase travel time for waves travelling between two selected points. Applying specific Fourier filters to the data and methodically selecting the points isolates waves which bounce through the solar interior several times to the far-side and back to the near-side. This multi-skip analysis relies on local conditions perturbing the phase travel time, and combinations of measurement configurations work to refine the signal \citep{fartd1,fartd2}.

The previously mentioned works, generally, have only focused on the detection of existing active regions on the far-side surface, though the same methods have also been applied to the pre-emergence detection of near-side active regions. Investigation of helioseismic holographic techniques go as far back as the 1990's, with notable developments made by \cite{Lindsey} in detecting sub-surface flows prior to active region emergence. A detailed study of emerging active regions using the holographic technique was performed by \cite{Birchp2} and statistically significant differences in the mean travel time deviation were observed. These mean travel time deviations are used by the authors to estimate the strength of converging flows surrounding the rising magnetic flux. Additionally, more recent tests on models of sunspots using holography, as in \cite{magtherm}, have attempted to separate the effects of thermal and magnetic contributions to acoustic travel times.

Some early examples of studies performed with the time-distance technique are \cite{earlyKoso} and \cite{Sunspot}, where the theoretical framework for time-distance helioseismology as well as detailed inversion procedures are provided. The authors apply their inversion procedure to an emerging active region observed by the Michelson Doppler Imager (MDI), where they successfully identify subsurface structures and estimate the rate of emergence of the magnetic flux. \cite{Kholikov} also studies pre-emergence active regions, some of which are the same regions as in \cite{earlyKoso}; it is found here that most travel time maps of active regions show a 10-15 second perturbation around the eventual location of the region on the solar surface. Further discussion of helioseismic methods and results are given in \cite{Gizon}, as well as a summary of efforts to detect sub-surface sound speed perturbations.

Of particular note are the results of \cite{Stathis2012} and \cite{Stathis2013}; the first work establishes a measurement procedure which produces a reported SNR of 4.2-4.9, while the second work further explains the procedure and examines other helioseismic signatures, such as frequency shifts, which may be indicative of emerging magnetic flux. The distinguishing features of this method are the division of annuli into several pairs of opposing arcs, as well as averaging of cross-correlation functions over these arcs and over groups of annuli. Using their method, the authors report a positive detection of emerging magnetic flux as well as estimates of the rate of flux emergence.

In this work, we use a modified version of the time-distance method developed by \cite{Stathis2012} to detect phase travel time shifts prior to the emergence of several active regions. We first test the method on numerical simulations to determine the sensitivity to sub-surface sound-speed perturbations. We then examine two active regions already studied by \cite{Stathis_nature}, NOAA AR 10488 and AR 07978, with data obtained by MDI. We also examine three more recent active regions, AR 11158, AR 12257, and AR 12772 with data obtained by the Helioseismic and Magnetic Imager (HMI). The results of the active region analysis are compared with a similar analysis of quiet sun regions, where the ambient magnetic field is relatively uniform and no significant features develop over the duration of data collection. We discuss characteristics of noise in the phase travel time measurements of these quiet sun regions, and suggest criteria to distinguish between true positive and false positive emerging flux detection.

\section{Methods and Data}

Half of the observational data used are MDI Dopplergrams covering an area of 30 by 30 degrees, with a resolution of 0.12 degrees per pixel, and 480 frames with 60 second cadence. The remaining half of observational data used are HMI Dopplergrams, covering identical spatial extent and duration, with 45 second cadence and a downgraded resolution of 0.12 degrees per pixel. The high resolution of HMI - 0.06 degrees per pixel - does not noticeably improve phase travel time measurements, and only increases computation time.  Both the HMI and MDI Dopplergrams are remapped to a Postel's projection and tracked at the Carrington rotation rate. We use simulated data from two separate 3D hydrodynamic models for testing, the first is developed by \cite{Hartlep_model} (referred to as Model 1) and the second developed by \cite{Andrey} (referred to as Model 2). Data from model 1 has a spatial resolution of 0.7 degrees per pixel and 60 second cadence, and data from model 2 has the same resolution and cadence as MDI.

\subsection{Measuring Shifts in Phase Travel Time}

\begin{figure}[h]
\begin{center}
    \includegraphics[width=0.5\linewidth]{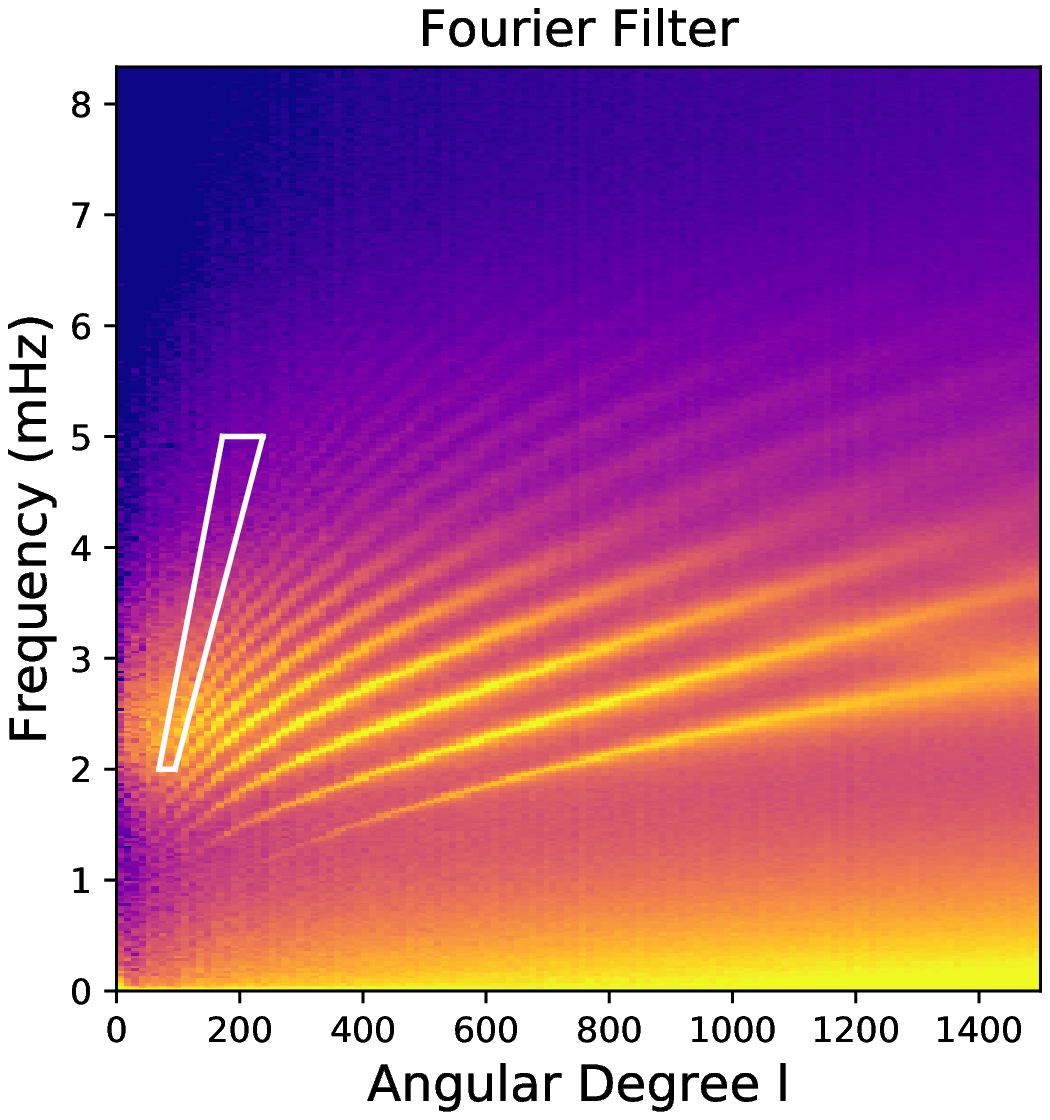}
    \caption{Power spectrum of a typical datacube used to compute phase travel time maps. The upper and lower horizontal white lines correspond to the edges of the flat-top filter at 2 and 5 mHz. The diagonal white lines form the center for the upper and lower phase speed filters. A Gaussian phase speed filter of width 8.7 km/s is applied to data outside the white box.}
    \label{powerspec}
\end{center}
\end{figure}

To begin, the data is treated with a dual Fourier filter; the first is a flat-top frequency filter which preserves oscillations only in the 2 to 5 mHz range, the resonant portion of the acoustic power spectrum. The second Fourier filter is a piece-wise Gaussian in phase speed, $v_{p}=\omega/k$, and the parameters are selected to isolate oscillations with a turning point between 40 and 70 Mm beneath the photosphere. The Standard Solar Model \citep{ssmodel} is used to estimate the sound speed at these depths, which are $v_{p1}=92$ km/s at 40 Mm and $v_{p2}=127$ km/s at 70 Mm. Oscillations with phase speeds slower than $v_{p1}$ are filtered with a Gaussian centered at $v_{p1}$ with a width of $\delta v = 8.7$ km/s. Oscillations between between $v_{p1}$ and $v_{p2}$ are untreated, and the oscillations above $v_{p2}$ are filtered with a Gaussian centered at $v_{p2}$ with the same width as the lower filter. In other words, the filters used are
\[
F_{1}(\nu) = \left\lbrace\begin{array}{cc}
    1 & 2 mHz \le \nu \le 5 mHz\\
    0 & otherwise
\end{array}\right.
\]
and
\[
F_{2}(v_p) = \left\lbrace\begin{array}{cc}
    \exp\left[-\dfrac{\left(v_p-v_{p1}\right)^2}{2{\delta v}^2}\right] & v_{p}<v_{p1} \\
    1 & v_{p1}\le v_{p} \le v_{p2} \\
    \exp\left[-\dfrac{\left(v_p-v_{p2}\right)^2}{2{\delta v}^2}\right] & v_{p2} < v_{p}  
\end{array}\right. ,
\]
and Figure \ref{powerspec} shows a typical power spectrum with the bounds of the Fourier filters highlighted in white.

Now that the data has been filtered to isolate oscillations with appropriate turning points, the cross-correlations are computed as in \cite{xcor}. Each pixel in the image has its corresponding spherical coordinates determined based on the heliographic location of the image's center pixel, and concentric rings centered on a given pixel are selected with radii equal to half the horizontal travel distance of the desired oscillations. The radius of the first ring is 4.56 degrees, corresponding to oscillations with the turning point at 40 Mm, and the radii increase in increments of 0.12 degrees - MDI spatial resolution - up to 8.16 degrees, corresponding to oscillations with the turning point at 70 Mm. Once the rings have been identified, 5 different configurations are used to compute the cross-correlation: the first configuration subdivides the ring into 3 opposing pairs of arcs, the second into 4 opposing pairs, and so on until the fifth configuration with 7 opposing pairs. Signals within these arcs are spatially averaged and used to compute a point-to-point cross-correlation for the selected turning point depth, arc configuration, and spatial position.

The cross-correlations for positive and negative time lags are computed separately and are fit to a Gabor wavelet \citep{Gabor_nigam} of the form
\[
\Psi(\tau) = A\cos\left(\omega_0(\tau-\tau_{ph})\right)\exp\left[-\frac{1}{2}\gamma^2(\tau-\tau_{g})^2\right],
\]
where $\tau$ is the time lag, $\tau_{ph}$ is the phase travel time, $\tau_{g}$ is the group travel time, $\omega_{0}$ is the dominant frequency of oscillation, and $\gamma$ is the decay rate of the Gaussian envelope. The cross-correlations are averaged over spatial position and are then fit using a non-linear least squares algorithm so that the phase travel time is a function of annulus radius and number of segments used to compute the cross-correlations. The initial values used for phase travel time, group travel time, and frequency are their theoretical values for the given horizontal travel distance. The cross-correlations are then shifted in time so that the mean phase travel times for each annulus radius are the same, allowing the cross-correlations to be averaged over distance. The cross-correlation for each pixel is then fit again to obtain the phase travel time, as a function of number of segments used. The phase travel time maps are then averaged to produce a final travel time map.

While we use the same cross-correlation procedure as in \cite{Stathis2012}, we choose to use a different Gabor wavelet fitting procedure with the differences being the following. First, the authors average their cross-correlations over the different cases of number of arcs used, where we keep these cross-correlations separate. The phase travel times differ slightly between the widest and narrowest arcs, presumably because of waves which scatter at the turning point leading to a "bent" ray path and slightly longer travel time. Narrower arcs more closely approximate the traditional, straight ray path while the wider arcs are more likely to capture a perturbed ray path, causing the disparity in phase travel time. Second, \cite{Stathis2012} use Quiet Sun values for their pixel-by-pixel fitting, where we only use these values to perform an initial fit on the spatially averaged cross-correlation which serves as an initial guess for our pixel-by-pixel fit. Finally, the authors use the Quiet Sun phase travel time to determine the shift for the cross-correlations to be averaged over distance, while we use the phase travel time from the spatially-averaged cross-correlation for each distance to determine the appropriate shift. These procedures produce identical shifts if the difference between the phase travel time for Active Regions and Quiet Sun regions is less than the observational cadence, but this cannot be guaranteed for ray paths which are especially long or encounter significant perturbations.

It can be shown - as in \cite{Kosovichev_tomo} - that the first order perturbation to phase travel time due to a change in sound speed is
\begin{equation}\label{delt}
\delta\tau_{mean}=\int_{\Gamma}\dfrac{\delta c}{c}U ds,
\end{equation}
where $\Gamma$ is the acoustic ray path, $\delta c$ is the perturbation to the sound speed $c$, and $U=\omega/k$ is the phase slowness. The mean perturbation to phase travel time, $\delta\tau_{mean}$, is computed as the simple average of the travel time perturbations from the positive and negative time lags, $\delta\tau_{\pm}=\tau_{\pm}-\langle\tau_{\pm}\rangle$. The reference phase travel time for the positive or negative time lag, $\langle \tau_{\pm}\rangle$, is obtained by taking a spatial average of the phase travel time maps. For perturbations which occupy a small portion of the map, this average should be approximately equal to the quiet Sun phase travel time.

\section{Results}
\subsection{Testing on Numerical Simulations: Model 1}

We begin with testing the time-distance method on numerical data from Model 1 \citep{Hartlep_model}, with two 5\% reductions in sound speed of varying spatial extent. Both perturbations are functions of depth (z) and angular distance ($\alpha$) from the horizontal center,
\[
\delta c = \dfrac{A}{4}\left(1+\cos\left(\pi\dfrac{z-z_{0}}{\delta z}\right)\right)\left(1+\cos\left(\pi\dfrac{\alpha}{\delta\alpha}\right)\right).
\]

\begin{figure}
\begin{center}
    \includegraphics[]{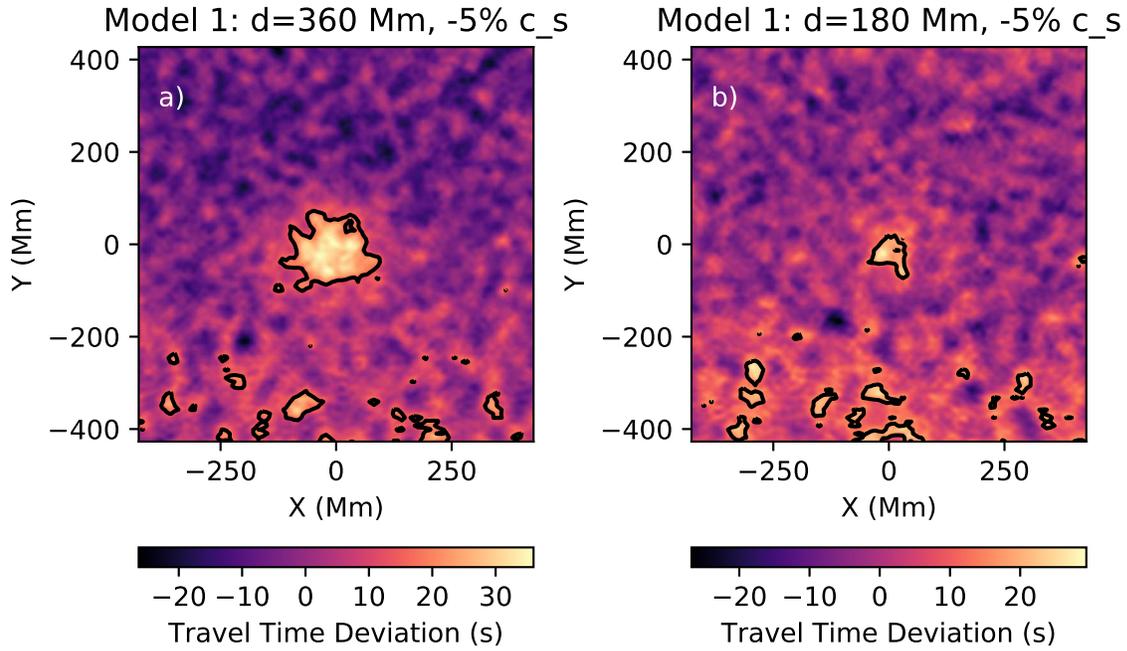}
    \caption{Measurements of phase travel time deviation for the sound speed perturbations in Model 1. a) Measurements for the perturbation with horizontal width of 360 Mm; b) Measurements for the perturbation with horizontal width of 180 Mm. The solid black line denotes the 2$\sigma$ threshold of 8.20 seconds derived from the unperturbed regions of the model.}
    \label{Hcube}
\end{center}
\end{figure}

The large perturbation has a horizontal width ($2\delta\alpha$) of 360 Mm and the smaller perturbation has a horizontal width ($2\delta\alpha$) of 180 Mm; both have a radius ($\delta z$) of 20 Mm in the vertical direction and are centered 50 Mm ($z_{0}$) beneath the model's surface. The large perturbation (Figure \ref{Hcube}a) is very well-resolved, with the strongest signal covering a circular area roughly 200 Mm in diameter. The smaller perturbation (Figure \ref{Hcube}b) is distinguishable from the background noise, but the spatial features are less clear than in the larger perturbation. Additionally, the magnitude of the travel time shift is reduced in the smaller perturbation, though this is expected as acoustic waves will travel less distance within the perturbed region which reduces the shift in travel time.

To highlight the region where the phase travel time is significantly perturbed, we set a 2$\sigma$ threshold of 8.20 seconds and overlay this contour with a black line in Figures \ref{Hcube}a and \ref{Hcube}b. This threshold is obtained by excluding data within 200 Mm of the map center which leaves only the unperturbed regions, and the standard deviation is computed from the remaining phase travel times. The shape of the contour in both cases differs slightly from the circularly-symmetric perturbations, indicating that detected spatial features may not necessarily coincide with \textit{in-situ} conditions. This is further supported by the discrepancy in Figure \ref{Hcube}a between the perturbation width of 360 Mm and a width of 200 Mm for the detected feature. Analysis of a different data set from Model 1 in \cite{Stathis2013} also yielded such a discrepancy. The contouring also highlights some anomalous travel times in the lower parts of Figures \ref{Hcube}a and \ref{Hcube}b for which we have no certain explanation. While it is expected that some pixels not associated with the perturbation will still exceed the threshold, both figures show a definite latitudinal dependence of the phase travel time deviation. The grid of Model 1 is uniform in latitude, so these anomalies may arise from the changing spatial resolution closer to the poles.

\subsection{Testing on Numerical Simulations: Model 2}

\begin{figure}
\begin{center}
    \includegraphics[width=0.75\linewidth]{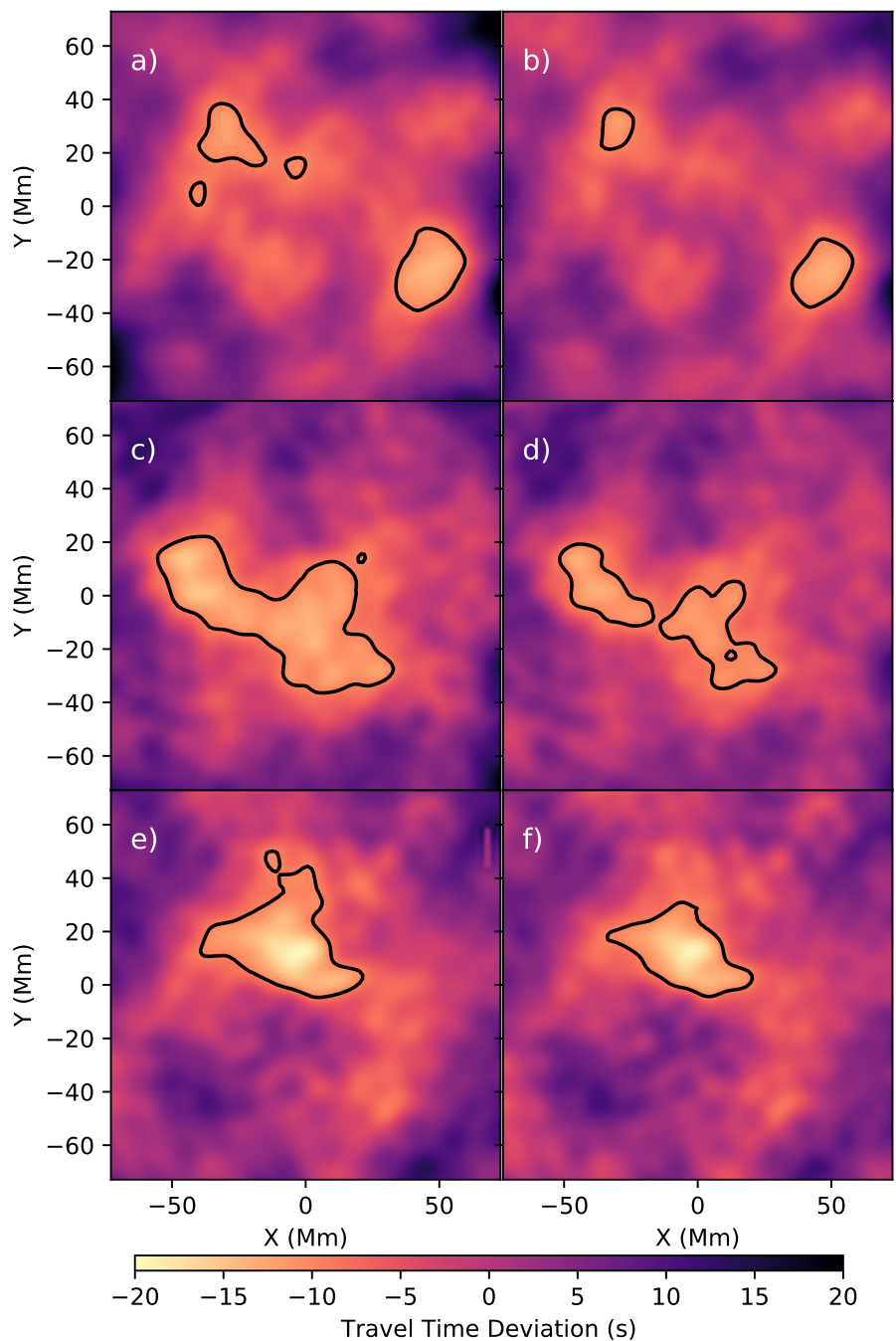}
    \caption{Travel time shifts measured in Model 2 from a 5\% increase in sound speed with a vertical radius of 20 Mm, at a depth of 40 Mm (a,b), 50 Mm (c,d), and 60 Mm (e,f). The horizontal radius is 50 Mm for (a,c,e) and 20 Mm for (b,d,f). The solid black line denotes the 2$\sigma$ threshold of 9.58 seconds derived from empty model runs.}
    \label{Andrey}
\end{center}
\end{figure}

\begin{figure}
\begin{center}
    \includegraphics[width=0.75\linewidth]{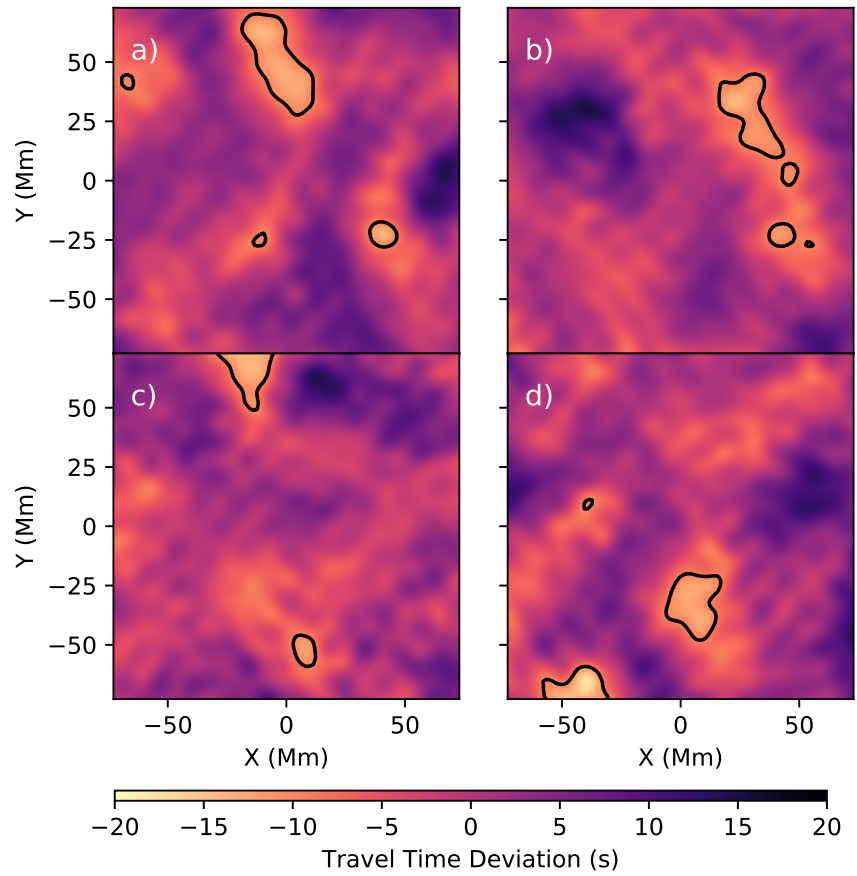}
    \caption{Travel time shifts measured in Model 2 for the unperturbed case. These maps are used to determine the $2\sigma$ threshold (shown here by the black contours) of Figures \ref{Andrey} and \ref{comp}.}
    \label{blank}
\end{center}
\end{figure}

We perform tests using Model 2 \citep{Andrey} on two characteristics of the time-distance method: dependence of sensitivity on depth and perturbation width, and the accuracy of the travel time measurements. Figure \ref{Andrey} shows six travel time maps of a perturbations with a radius in the vertical direction of 20 Mm. The left column contains perturbations with a horizontal radius of 50 Mm and the right column contains perturbations with a horizontal radius of 20 Mm. From top to bottom, the depths of the perturbations are 40, 50, and 60 Mm. Each perturbation has an amplitude of 5\% of the background sound speed and is Gaussian in all directions. As with Model 1, we use the standard deviation of unperturbed regions to set a 2$\sigma$ threshold to highlight strongly perturbed regions. Four unperturbed cases are used for this computation, and the resulting travel time maps are shown in Figure \ref{blank}. The 2$\sigma$ contour is also applied to these maps.

The most surprising feature of the perturbed travel time maps is the lack of spatial information; the perturbations are Gaussian in the (x,y)-plane which should imply a circular travel time shift. We see instead that the spatial structure is convolved with the background noise, and the travel time shift contribution added to underlying inhomogeneities. This is particularly true for the perturbation placed at a depth of 40 Mm, which is the upper limit of the vertical region that we sample. Furthermore, there is very little change in the shape of the travel time measurements between the 20 and 50 Mm perturbations at 50 and 60 Mm. Some similarities are expected as the oscillation excitation functions are the same, but the only distinguishing feature between the two are the magnitude of the travel time shifts. The larger sound speed perturbation produces travel time shifts which are two to three seconds greater than the smaller perturbation. The travel time shifts also increase in magnitude with the depth of the perturbations; since the sound speed perturbation is proportional to the local sound speed, we should expect the phase travel time to vary proportionally as well.

\begin{figure}
\begin{center}
    \includegraphics[]{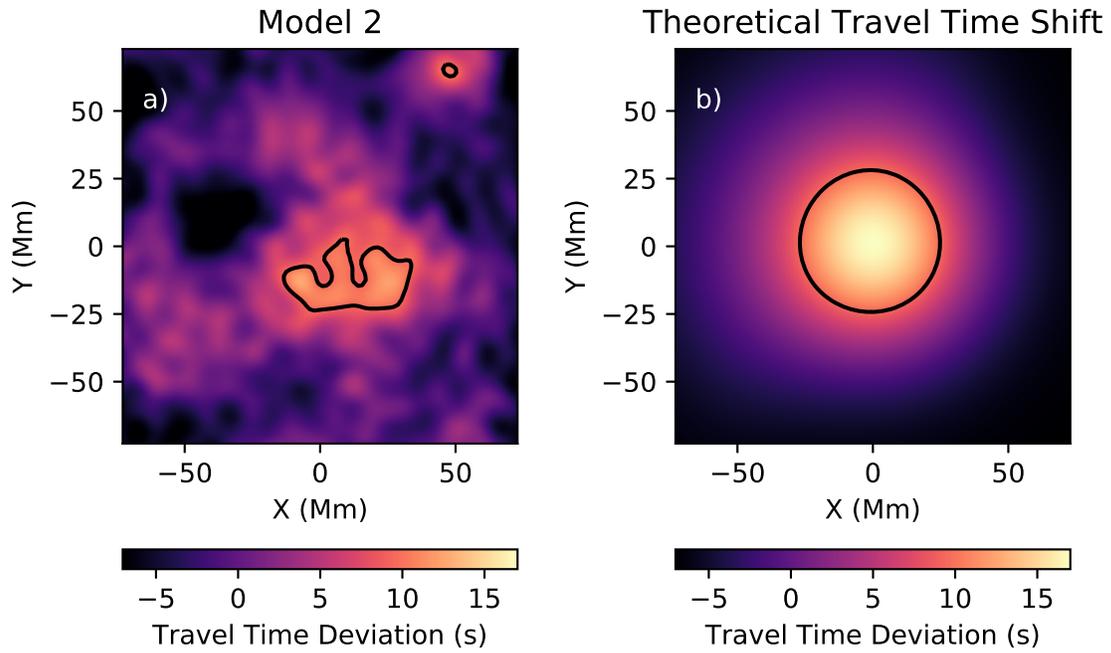}
    \caption{Comparison between the measured (a) and theoretical (b) travel time shifts for a 5\% reduction in sound speed. The scale is adjusted to align with the maximum and minimum of the theoretical travel time shifts. The solid black line denotes the 2$\sigma$ threshold of 9.58 seconds derived from empty model runs.}
    \label{comp}
\end{center}
\end{figure}

To test the accuracy of the travel time measurements, we place a 5\% reduction in sound speed at 55 Mm below the model's surface, with a radius of 20 Mm in all directions. The theoretical travel time shift is computed using equation \ref{delt} and the standard solar model \citep{ssmodel} is used as the background mesh. The travel time shifts are sampled similarly to the time-distance measurement method, with signal averaged between several arc pairs, though we select only one ray per arc pair. This is in contrast to the cross-correlation procedure, where the Doppler signal is averaged over a given arc, which is analogous to averaging over many rays in the ray-path computation. The theoretical travel time map has the mean subtracted as is done with the measurements of the perturbation, and the results for both are shown in Figure \ref{comp}. The area in the measured travel time map with the greatest amplitude roughly corresponds to the full-width half-maximum of the theoretical signal. Interestingly, the measured travel time shifts are several seconds shorter than what would be expected, though as previously discussed, the spatial structure of the perturbation is convolved with the background noise. While this difference can be explained by the noise level (4.79 seconds), it's expected that this noise would contribute equally to an increase in the travel time in at least some pixels. However, we observe far more reductions in the travel time magnitude than increases.

\subsection{Detection of Emerging Flux and Comparison with Quiet Sun Measurements}

We apply the time-distance method to five active regions using Dopplergrams from HMI and MDI, and first examine NOAA ARs 10488 and 7978 which were previously studied by \cite{Stathis_nature}, who measured strong travel time shifts several hours prior to the emergence of the active regions. To quantitatively evaluate the overall strength of the travel time shifts, we define the perturbation index to be the sum of negative travel time shifts with amplitude two times greater than the standard deviation of the data series, multiplied by the area these travel time shifts occupy. For example, a feature measuring 50 Mm$^2$ with a magnitude of 10 seconds will have a perturbation index of 500 s Mm$^2$. Thus the perturbation index will increase if either the area or the magnitude of a travel time shift increases.

\begin{figure}
\begin{center}
    \includegraphics[]{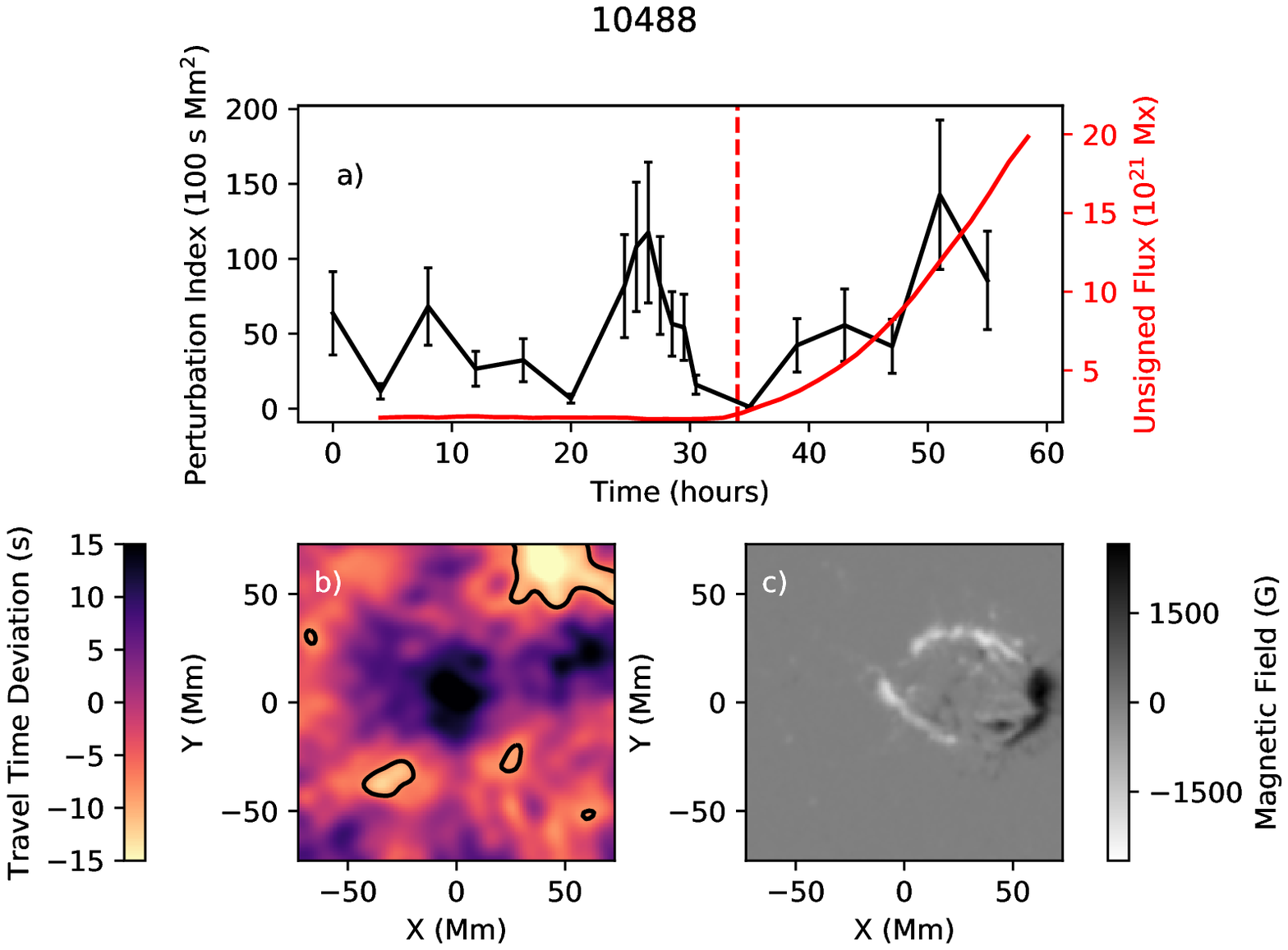}
    \caption{Analysis of AR 10488. a) Perturbation index over time (black) and the unsigned magnetic flux (red); the vertical red dashed line indicates the approximate time of emergence. b) Travel time map at the time of peak perturbation index; the regions which meet the 2$\sigma$ threshold are bounded by the solid dashed line. c) Magnetogram of the active region at the end of the data series.}
    \label{10488}
\end{center}
\end{figure}

\begin{figure}
\begin{center}
    \includegraphics[]{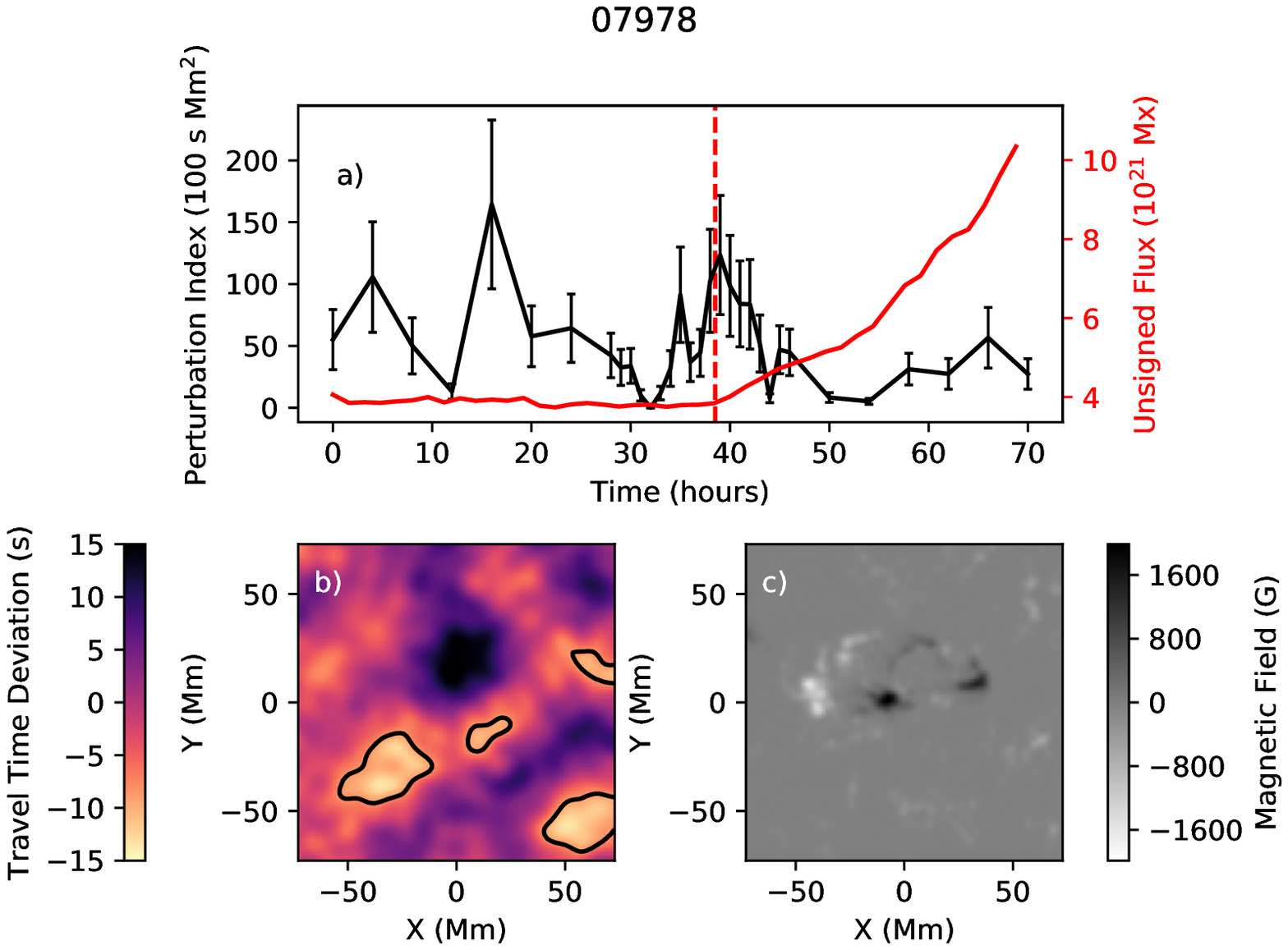}
    \caption{Analysis of AR 7978. a) Perturbation index over time (black) and the unsigned magnetic flux (red); the vertical red dashed line indicates the approximate time of emergence. b) Travel time map at the time of peak perturbation index; the regions which meet the 2$\sigma$ threshold are bounded by the solid black line. c) Magnetogram of the active region at the end of the data series.}
    \label{07978}
\end{center}
\end{figure}

AR 10488 shows a small increase in the perturbation index 25 hours before emergence and stronger increase five hours before emergence, as shown in Figure \ref{10488}a. A large peak in the perturbation index appears roughly 25 hours after AR 10488 begins emerging; extended observations by \cite{Stathis_nature} show that the magnetic flux rate decreases one day after emergence, before abruptly increasing. We interpret this strong peak as additional flux which strengthens the existing magnetic field. The perturbation index of AR 7978, shown in Figure \ref{07978}a is slightly more noisy than that of AR 10488, and shows a strong peak 20 hours before emergence. The magnetic flux of the active region increases slowly for about 15 hours after emerging, and then experiences a strong increase in the remaining 15 hours of observation. A moderate peak centered around the time of emergence precedes the change in the flux rate by 13 to 15 hours.

\begin{figure}
\begin{center}
    \includegraphics[]{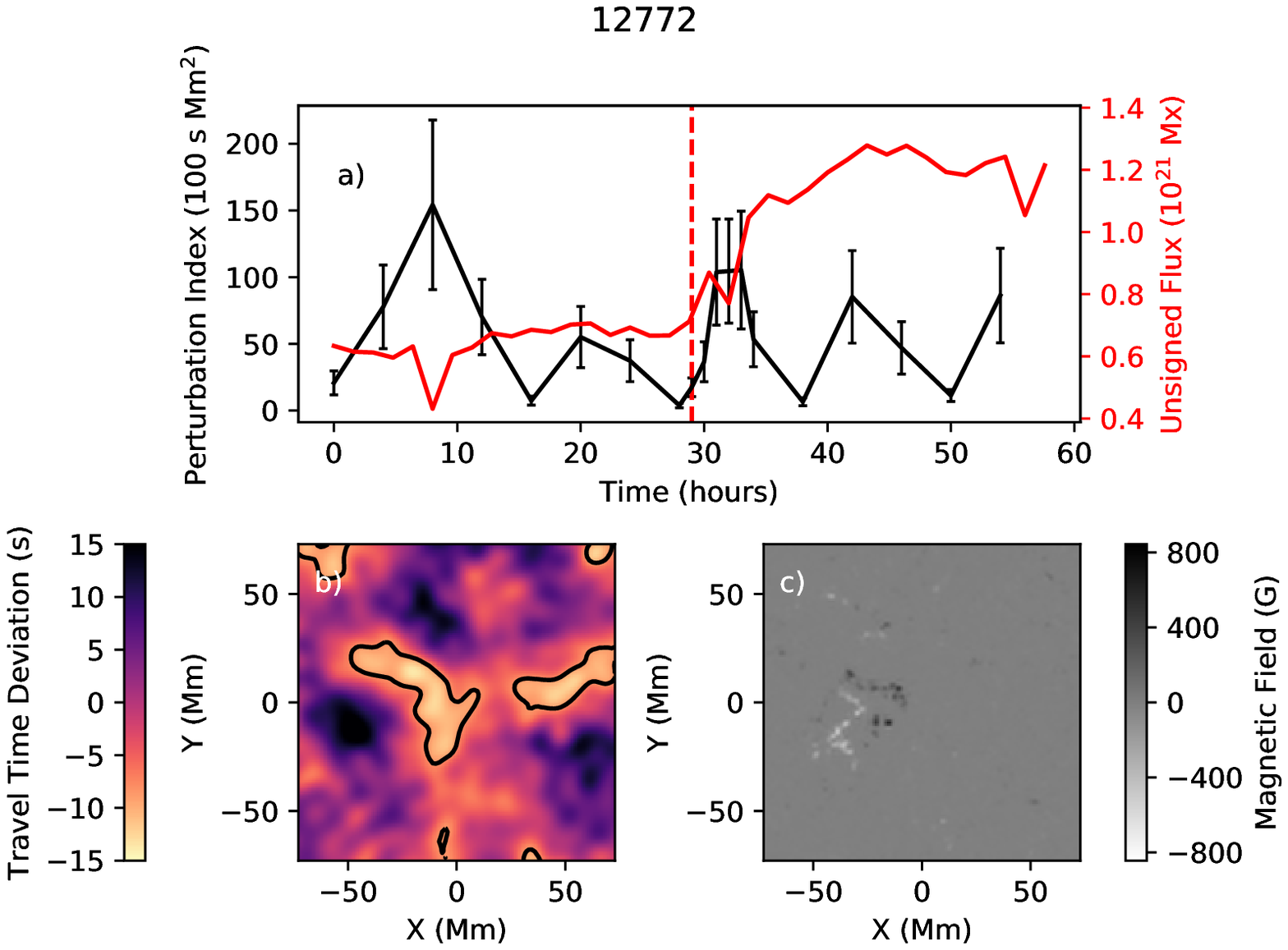}
    \caption{Analysis of AR 12772. a) Perturbation index over time (black) and the unsigned magnetic flux (red); the vertical red dashed line indicates the approximate time of emergence. b) Travel time map at the time of peak perturbation index; the regions which meet the 2$\sigma$ threshold are bounded by the solid black line. c) Magnetogram of the active region at the end of the data series.}
    \label{12772}
\end{center}
\end{figure}

\begin{figure}
\begin{center}
    \includegraphics[]{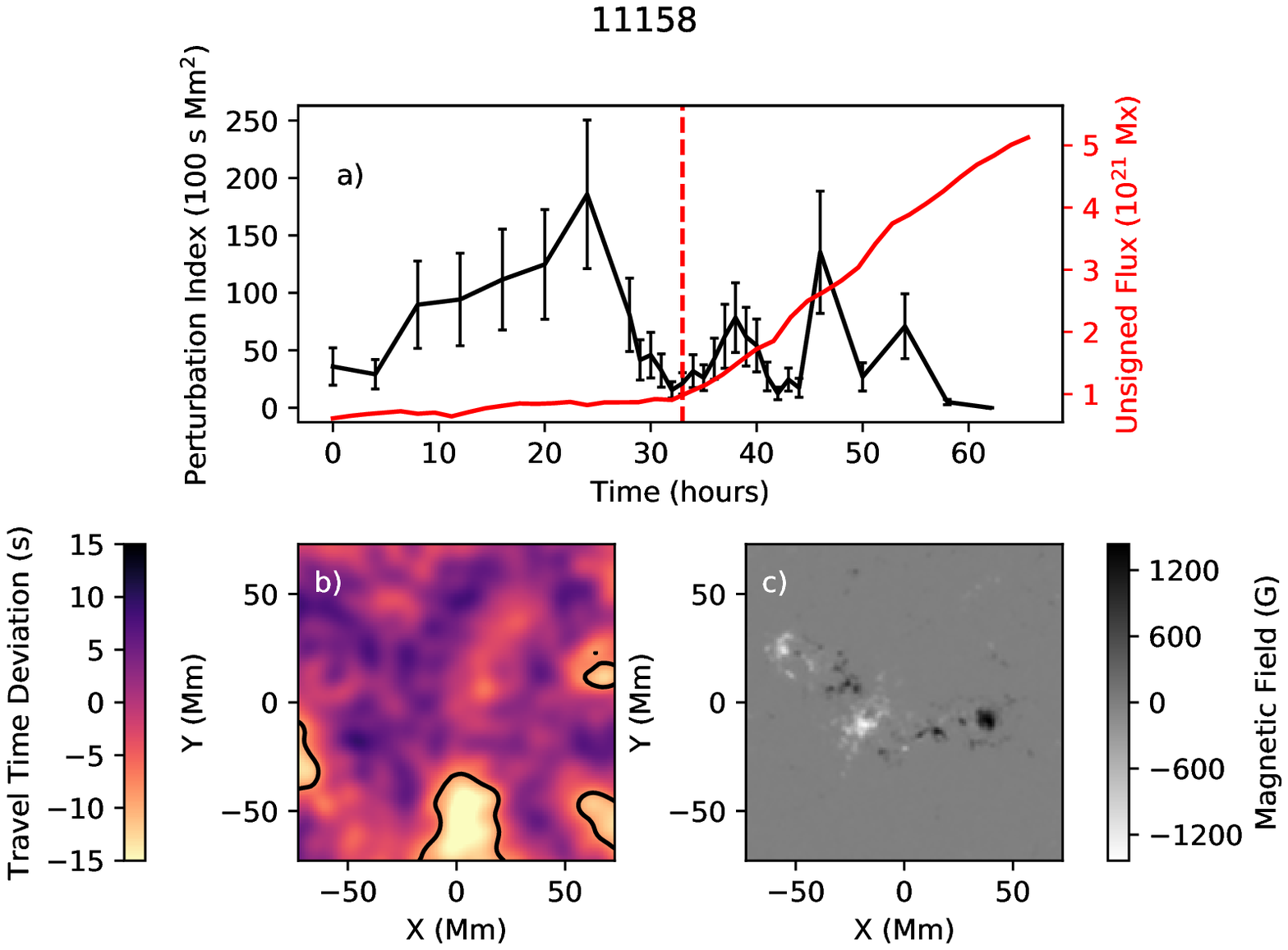}
    \caption{Analysis of AR 11158. a) Perturbation index over time (black) and the unsigned magnetic flux (red); the vertical red dashed line indicates the approximate time of emergence. b) Travel time map at the time of peak perturbation index; the regions which meet the 2$\sigma$ threshold are bounded by the solid black line. c) Magnetogram of the active region at the end of the data series.}
    \label{11158}
\end{center}
\end{figure}

\begin{figure}
\begin{center}
    \includegraphics[]{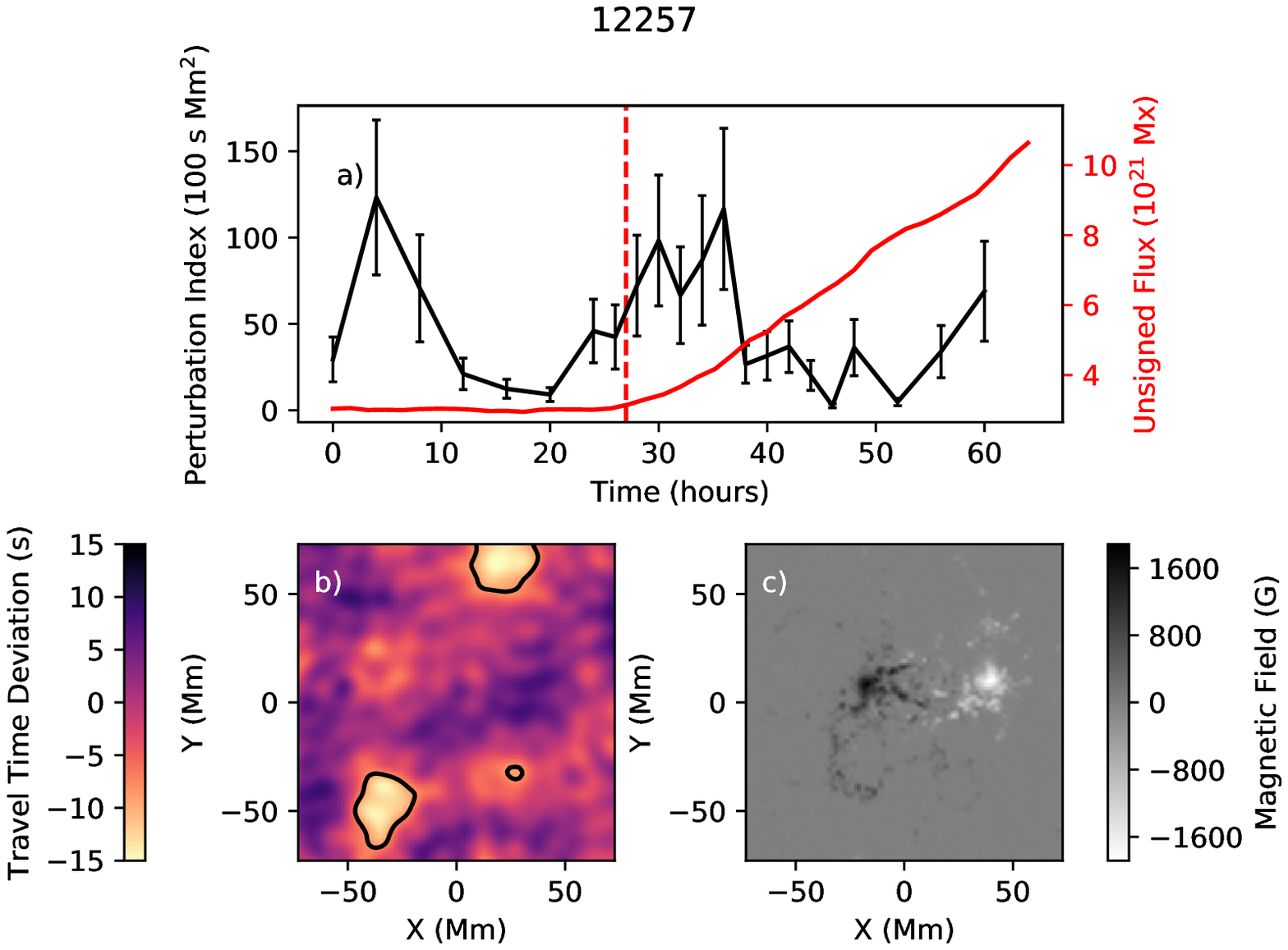}
    \caption{Analysis of AR 12257. a) Perturbation index over time (black) and the unsigned magnetic flux (red); the vertical red dashed line indicates the approximate time of emergence. b) Travel time map at the time of peak perturbation index; the regions which meet the 2$\sigma$ threshold are bounded by the solid black line. c) Magnetogram of the active region at the end of the data series.}
    \label{12257}
\end{center}
\end{figure}

AR 12772 is a relatively weak active region, with an unsigned magnetic field not exceeding 850 Gauss. We are still able to detect a significant peak in the perturbation index (Figure \ref{12772}a) 20 hours before the region emerges. In contrast, AR 11158 is a fairly strong active region, with the magnetic field exceeding 1000 Gauss at the end of our observations. The configuration of the field also has a more significant bipolar structure than AR 12772. For this region, we see a gradual rise in the perturbation index (Figure \ref{11158}a) 25 hours prior to emergence which peaks around 10 hours before the unsigned flux begins increasing. Additionally, there are some smaller changes following initial emergence which precede sharp changes in the unsigned magnetic flux. Finally, we see a moderate increase in the perturbation index of AR 12257 (Figure \ref{12257}a) 20 hours before the region emerges, and a slightly smaller increase centered around the region's emergence. This second increase is more representative of expected emergence behavior: a gradual rise in the perturbation index, followed by a sharp decline as the concentration of magnetic flux rises out of focus.

\begin{figure}
    \centering
    \includegraphics[width=\linewidth]{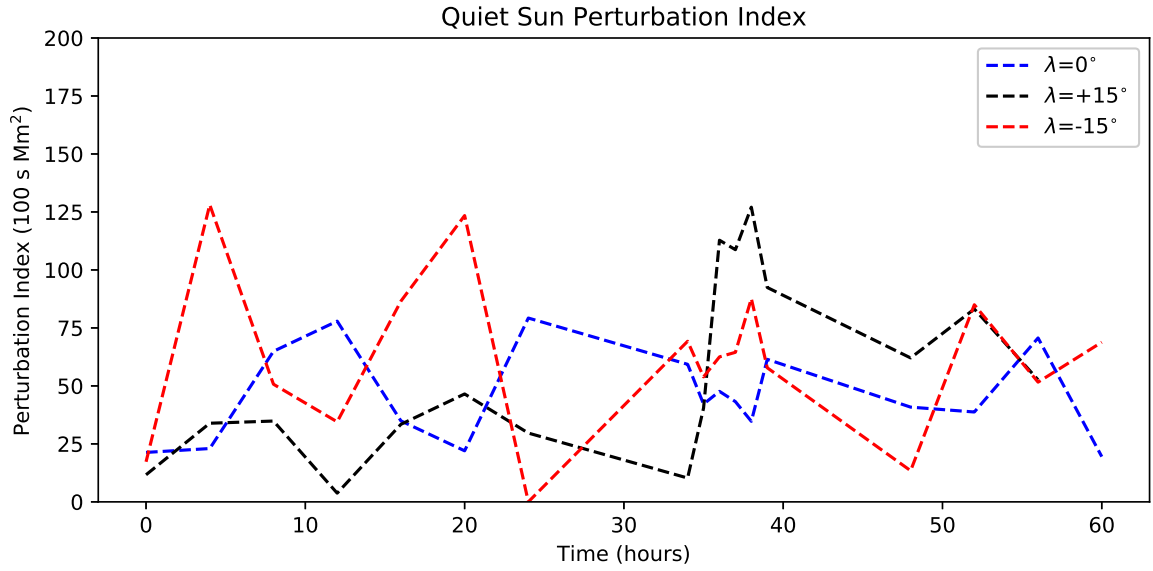}
    \caption{Perturbation index of 3 Quiet Sun regions with heliographic latitude +15$^{\circ}$ (black), 0$^{\circ}$ (blue), and -15$^{\circ}$ (red), and equal longitude.}
    \label{QS}
\end{figure}

In contrast to the perturbation index of the active regions, the Quiet Sun regions show no gradual increases of the perturbation index, as observed in some active regions, over 60 hours of observation (Figure \ref{QS}). The perturbation index of the Quiet Sun region at a latitude of +15$^{\circ}$ does show a rapid increase at 38 hours followed by a gradual decline, though this isn't observed in the other Quiet Sun regions or any of the active regions. We expect, and observe in AR's 11158 and 12257, an opposite trend where there is a gradual increase in the perturbation index followed by a sharp decline. This can be interpreted as a concentration of magnetic flux just emerging into focus at 70 Mm, slightly perturbing the phase travel time. As the flux rises, it encounters more acoustic rays in our observed depths of 40-70 Mm, further perturbing the measured phase travel time. The perturbation index will peak once the lower extent of the magnetic flux passes through a depth of 70 Mm, then decrease quicker than the initial rise. This asymmetrical development comes from the uneven sensitivity of acoustic rays which we use to probe the interior. As evidenced in Figure \ref{Andrey}, deeper perturbations can cause a stronger phase travel time shift as compared to a perturbation with equal magnitude at a shallower depth. Since acoustic rays which penetrate to 70 Mm have less curvature than the rays which penetrate to 40 Mm, a wave packet will travel more distance in the region close to the turning point, making it more sensitive to travel time shifts.

\section{Discussion and Conclusion}

Aside from the rate at which the perturbation index rises, we should also take into account the magnitude of the peak when evaluating whether or not an active region will emerge. The perturbation index of Quiet Sun regions can occasionally reach values close to 13000 s Mm$^2$, though it is more likely that it will fluctuate around a mean of 4000 to 6000 s Mm$^2$. If we choose the strict criteria that a pre-emergence active region can only be identified by exceeding any Quiet Sun perturbation index values (i.e exceed 13000 s Mm$^2$), then a successful identification can be made in three of the five active regions studied - not including the secondary emergence signature detected in AR 10488.

While the perturbation index is a useful tool in looking for pre-emergence signatures, it is difficult to to determine the uncertainty associated with it, and this is especially true for observational data. Our procedure involves several non-linear operations - such as the Gabor wavelet fitting - which confounds the uncertainties of the original measurements. It could be measured from the previously discussed acoustic models without any sound-speed perturbations, but these models lack true convection and other important processes which may distort the phase travel time. We decide to compute the uncertainty using the Quiet Sun measurements, under the assumption that there are no significant inhomogeneities present. We compute the RMS of the phase travel time deviation for each frame in the 60-hour series of the three latitudes studied, 4.307 seconds, and use this as the noise level. The uncertainty in the perturbation index at each frame is then computed as the area of the detected perturbation multiplied by the noise level.

Unfortunately, detailed spatial features cannot be observed using the time-distance method, as the average wavelength of oscillations which reach our observation depths is around 3 Mm. For this reason, we use MDI resolution (0.12 degrees per pixel) in our HMI observations, as the travel time will have minimal variation between pixels at a resolution of 0.06 degrees per pixel. This also speeds up the computation for HMI phase travel time maps compared to using the native resolution. While we cannot measure significantly detailed spatial features, the placement of positive and negative travel time shifts relative to the eventual positions of active regions is interesting. At the time of peak emergence, there are areas of positive travel time shift coincident with the eventual location on the surface of magnetic flux. The negative travel time shifts, which are used to compute the perturbation index, are adjacent to the eventual concentrations of magnetic flux on the surface.

In earlier stages of emergence, we measure negative travel time shifts coincident with the surface magnetic flux, as in the work of \cite{Stathis_nature}, though the later, adjacent shifts are stronger in our observations. In response to the results of \cite{Stathis_nature}, \cite{Braun_comment} comment that their signatures could not be detected using a different method, and that the results are inconsistent with models of emerging flux. In response, \cite{Ilonidis_response} point to differences in the measurement techniques, including their higher SNR, and offer some possible physical processes which may alter the measured phase travel times. One of the primary arguments made by \cite{Braun_comment} is that direct perturbations to the sound speed due to the magnetic field are two orders of magnitude smaller than perturbations due to flows induced by the magnetic field. \cite{Braun_comment} cite \cite{mag}, who find that the radial flows are strongest above the magnetic flux and azimuthal flows are strongest adjacent to the flux; the authors also state that the azimuthal flows tend to dominate.  The information provided by \cite{Braun_comment} and \cite{mag} may explain our observation of strong travel time shifts adjacent to the active regions prior to emergence,  and magnetic stresses and flow perturbations may contribute to the observed travel time shifts. It should also be noted that while sound-speed perturbations in simulations can produce travel time shifts similar to the detected pre-emergence signatures, this does not necessarily imply that those signatures are caused by a sound-speed perturbation.

In this work, we used a time-distance method based on the ray approximation to compute phase travel time differences. We first applied this method to two numerical models in order to evaluate the sensitivity and spatial resolution. We find that large perturbations to sound speed are easily detectable, with moderate spatial detail. Smaller perturbations can be detected with the loss of some spatial information, though detection is difficult if the perturbation is too small or weak.

We then applied the time-distance method to five active regions - two observed by MDI and three by HMI - as well as Quiet Sun regions where there was no significant enhancement of the magnetic field. Measuring the perturbation index, a proxy for strength of travel time shifts, we are able to definitively identify three of five active regions before emergence, as well as secondary development in an additional active region. We confirm the detection of AR's 10488 and 7978 by \cite{Stathis_nature}, and find that the spatial position of travel time shifts in all studied active regions support the results of \cite{Braun_comment} and \cite{mag}. 

There is still much work to be done in developing this time-distance method. While our results show the method is viable, the relationship between the magnitude and duration of travel time shifts to the strength and size of an active region are still unclear. In further studies, we will perform a broad statistical survey of historical active regions observed by MDI and HMI. It is suspected that temporal development of the perturbation index can disentangle the false positives in Quiet Sun measurements from true identification of emerging flux, and this statistical survey should be able to validate or invalidate this hypothesis.

\acknowledgements{This work was supported by NASA grants NNX14AB70G, 80NSSC19K0268, 80NSSC19K1436, 80NSSC20K0602, and 80NSSC20K1870. The authors would like to thank Dr. Thomas Hartlep for his publicly available solar acoustic models, and the NASA Advanced Supercomputing Division and the Stanford Solar Group for providing access to observational data and computing facilities. Special thanks to the Heliophysics Modeling and Simulation Group at NASA Ames Research Center for their advice and support.}

\bibliography{mybib}

\end{document}